# Pressure Effects on Superconducting Properties of a BiS$_2$-based Superconductor Bi$_2$(O,F)S$_2$

Tomoyuki Okada[1], Hiraku Ogino[1]*, Jun-ichi Shimoyama[1], Kohji Kishio[1], Nao Takeshita[2], Naoki Shirakawa[2], Akira Iyo[2] and Hiroshi Eisaki[2]

[1] *Department of Applied Chemistry, The University of Tokyo, 7-3-1 Hongo, Bunkyo-ku, Tokyo, 113-8656, Japan*
2. *National Institute of Advanced Industrial Science and Technology, 1-1-1 Umezono, Tsukuba, Ibaraki, 305-8658, Japan*

Pressure effects on a recently discovered BiS$_2$-based superconductor Bi$_2$(O,F)S$_2$ ($T_c$ = 5.1 K) were examined via two different methods; high pressure resistivity measurement and high pressure annealing. The effects of these two methods on the superconducting properties of Bi$_2$(O,F)S$_2$ were significantly different although in both methods hydrostatic pressure is applied to the sample by the cubic-anvil-type apparatus. In high pressure resistivity measurement, $T_c$ linearly decreased at the rate of -1.2 K GPa$^{-1}$. In contrast, the $T_c$ of 5.1 K is maintained after high pressure annealing under 2 GPa and 470°C of optimally doped sample despite significant change of lattice parameters. In addition, superconductivity was observed in fluorine-free Bi$_2$OS$_2$ after high pressure annealing. These results suggest that high pressure annealing would cause a unique effect on physical properties of layered compounds.

## 1. Introduction

Since the discovery of superconductivity in Bi$_4$O$_4$S$_3$[1], various layered superconductors with rock-salt type BiS$_2$ layer have been reported. BiS$_2$-based superconductors have varied blocking layers such as fluorite-type $RE$(O,F) layer ($RE$ = La[2], Ce[3], Pr[4] and Nd[5]) and (Sr,La)F layer[6]. Recently some related compounds with complex stacked structure, such as Eu$_3$Bi$_2$S$_4$F$_4$ ($T_c$ = 1.5 K) [7] and LaPbBiS$_3$O (non-superconductor)[8], has also been reported.

Some of the BiS$_2$-based superconductors show increase of $T_c$ by applying high pressures (HP). In HP resistivity measurement, La(O,F)BiS$_2$ shows sudden increase of $T_c$ at ~1 GPa from ~3 K to ~10 K[9] and then $T_c$ slowly decreases at the slope of -0.28 K GPa$^{-1}$ [10]. Structural analysis at room temperature suggests that structural phase transition from tetragonal ($P4/nmm$) to monoclinic ($P2_1/m$) structure would be responsible for the increase of $T_c$[10]. Other $RE$(O,F)BiS$_2$ ($RE$ = Ce ~ Nd)[9,11], (Sr,La)FBiS$_2$[12], and Eu$_3$Bi$_2$S$_4$F$_4$[13] also shows the jump-up of $T_c$ to 7~10 K at a certain pressure of 1~2 GPa.

Similar increase of $T_c$ is also obtained by "HP annealing". In HP annealing, the as-synthesized samples undergo heat treatment in HP synthesis apparatus and physical properties are measured under ambient pressure. In La(O,F)BiS$_2$, $T_c$ increase from ~3 K to 10.6 K after HP annealing at 2 GPa, 600°C[2]. Ce(O,F)BiS$_2$[14] and Pr(O,F)BiS$_2$[15] also show increase of $T_c$ by HP annealing. In HP annealing, crystal structure is kept at tetragonal and uniaxial contraction of the lattice along $c$-axis is attributed to the cause for the increase of $T_c$[15,16].

Recently, superconductivity in Bi$_2$(O,F)S$_2$ (i.e. Bi(O,F)BiS$_2$), which is isostructural with $RE$(O,F)BiS$_2$, is reported[17)18)]. $T_c$ varies with changing F-doping level in the underdoped region and reaches to 5.1 K in the optimally-doped sample, while it is impossible to obtain overdoped sample due to the solubility limit of F$^-$ for the O$^{2-}$ site[17]. In the present paper, pressure effects on Bi$_2$(O,F)S$_2$ were examined by HP resistivity measurement and HP annealing. The difference of these two methods in the variation of $T_c$ in Bi$_2$(O,F)S$_2$ is discussed.

## 2. Experimental

Undoped Bi$_2$OS$_2$ was synthesized via solid-state reaction at 300°C, ~100 h in an evacuated quartz ampule with several intermediate grindings. Bi$_2$(O,F)S$_2$ was synthesized by the topotactic reaction between undoped Bi$_2$OS$_2$ and XeF$_2$ at 400°C, 72 h in an evacuated quartz ampule according to the following chemical reaction.

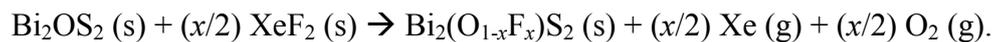
Bi$_2$OS$_2$ (s) + ($x$/2) XeF$_2$ (s) → Bi$_2$(O$_{1-x}$F$_x$)S$_2$ (s) + ($x$/2) Xe (g) + ($x$/2) O$_2$ (g).

Bi$_2$(O,F)S$_2$ samples with different F-content were separately prepared by changing the nominal amount of XeF$_2$ between ($x$/2) = 0.1~0.5. Note that actual composition of F is supposed to be smaller than the nominal $x$, while investigation of actual F content is difficult due to the small difference in X-ray scattering factors between O and F. Details of synthesis procedure are described in ref 17).

Resistivity measurements were performed by the four-probe method both in ambient pressure and under HP. Resistivity at ambient pressure is measured by Quantun Design Physical Property Measurement System (PPMS). In HP resistivity measurement, hydrostatic pressures up to 11 GPa were applied to the Teflon sample cell together with liquid pressure-transmitting medium Daphne oil 7474 in a cubic-anvil-type apparatus[19].

HP annealing was performed on the as-synthesized Bi$_2$OS$_2$ and Bi$_2$(O,F)S$_2$ at 300~470°C, 2 GPa for 1 h in a cubic-anvil-type apparatus. HP synthesis of Bi$_2$OS$_2$ was also performed from Bi$_2$O$_3$ (Furuuchi Chemical, 99.9%) and Bi$_2$S$_3$ (ALDRICH, 99%) powders at 300~550°C, 2~4 GPa. BN was used for the pressure media. The constituent phases of the samples were analyzed by X-ray diffraction (XRD) measurement (RIGAKU Ultima-IV) with Cu-$K_\alpha$ radiation generated at 40 kV and 40 mA. Silicon powder was used as an internal standard to determine the lattice parameters. Magnetization measurements were performed by a superconducting quantum interference device (SQUID) magnetometer (Quantum Design MPMS) and some measurements were performed down to 0.47 K using iHelium3[20].

## 3. Results

### 3.1 High pressure resistivity measurement on $Bi_2(O,F)S_2$

Hydrostatic high pressure resistivity measurement is performed for the optimally-doped $Bi_2(O,F)S_2$ polycrystalline bulk sample with $T_c$ = 5.1 K. The temperature dependence of resistivity under pressures of 0~11 GPa is shown in **Fig. 1**. $T_c$ monotonically decreases by application of hydrostatic pressure of 1~2 GPa. At 2 GPa, $T_c^{onset}$ was decreased to 2.7 K and zero resistivity was not observed above 2.5 K. The slope for the almost linear decrease of $T_c$ with applying pressure is calculated to be -1.2 K GPa$^{-1}$ as shown in **Fig. 2**. The superconducting transition remain sharp even under high pressures, with $\Delta T_c$ ~0.4 K, suggesting that high pressure is homogeneously applied to the bulk sample.

Under higher pressures than 2 GPa, superconductivity is disappeared. Instead, two other anomalies, marked $T_{a1}$ and $T_{a2}$ in **Fig. 1**, are observed. Origins of these anomalies are attributed to the superconductivity of Bi metal under high pressures. As shown in **Fig. 3(a)**, our sample contains small amount of Bi metal as an impurity phase. The ambient pressure phase of Bi metal, trigonal Bi(I), is not a superconductor. High pressures induce several steps of structural transitions to so-called Bi(II), Bi(III) and Bi(V) phases. Orthorhombic Bi(III) and bcc Bi(V) phases exhibit superconductivity and $T_c$s of Bi(III)[21] and Bi(V)[22] are very close to $T_{a1}$ and $T_{a2}$, respectively, as shown in **Fig. 2.**

The sudden increase of $T_c$ at a certain pressure observed in $RE$(O,F)BiS$_2$ ($RE$ = La~Nd)[9,11], (Sr,La)FBiS$_2$[12], and Eu$_3$Bi$_2$S$_4$F$_4$[13] was not observed in $Bi_2(O,F)S_2$. The decrease of $T_c$ with applying pressure is rather similar to the case of Bi$_4$O$_4$S$_3$[23] or Bi$_3$O$_2$S$_3$[24]. We note that crystal structure of Bi$_4$O$_4$S$_3$ and Bi$_3$O$_2$S$_3$ phase is still controversial[25], but both structure contains fluorite-type BiO layer in the blocking layer, as $Bi_2(O,F)S_2$ does.

### 3.2 High pressure annealing on $Bi_2(O,F)S_2$

High pressure (HP) annealing was performed for the as-synthesized $Bi_2(O,F)S_2$ with different F-content. Here we discuss the effects of HP annealing on two types of samples; underdoped sample with $T_c$ = 3.0 K, and optimally-doped sample with $T_c$ = 5.1 K.

**Figure 3(a)** shows powder XRD patterns of $Bi_2(O,F)S_2$ samples before and after HP annealing at 470°C and 2 GPa for 1 h. After HP annealing, constituent phases were almost unchanged and the peaks for $Bi_2(O,F)S_2$ were well indexed to the space group P4/*nmm* as well as the as-synthesized samples. *a*-axis length and *c*-axis length slightly increased and largely decreased, respectively, suggesting uniaxial contraction of the lattice along the *c*-axis as in the case of La(O,F)BiS$_2$[16] and Pr(O,F)BiS$_2$[15]. We note that the full width at half maximum (FWHM) of the XRD peaks were broadened after HP annealing, suggesting introduction of lattice strain.

The Zero-Field-Cooled (ZFC) magnetizations of these samples are shown in **Fig. 3(b)**. In the underdoped sample with $T_c$ = 3.0 K, $T_c$ increased to 4.5 K after HP annealing. Assuming that F-content of the sample is not changed by HP annealing, the uniaxial contraction of the lattice discussed above is supposed to be responsible for the increase of $T_c$.

In contrast, in the optimally-doped sample with $T_c$ = 5.1 K, $T_c$ almost unchanged even after HP annealing despite significant change of lattice parameters. The ZFC curve of the HP annealed sample seems to be somewhat shifted to lower temperatures. However, the onset of diamagnetic transition is kept at 5.1 K and perfect diamagnetism is achieved even at slightly higher temperature, indicating sharpening of the superconducting transition by HP annealing. The maintenance of $T_c$ at 5.1 K is rather similar to the case of Nd(O,F)BiS$_2$, which also maintains $T_c$ of ~5 K before and after HP annealing[26], than the case of RE(O,F)BiS$_2$ with RE = La~Pr where $T_c$ is increased by HP annealing[2,15,16]. Here it should be noted again that $T_c^{onset}$ of optimally-doped Bi$_2$(O,F)S$_2$ at 2 GPa was 2.7 K in HP resistivity measurement, significantly lower than that of the same sample annealed at 2 GPa. In HP annealing, high pressure is applied at high temperatures and physical properties are measured at ambient pressure. The difference of these two methods should not be dismissed.

3.3 *Superconductivity of high pressure annealed / synthesized Bi$_2$OS$_2$*

HP annealing is also performed for undoped Bi$_2$OS$_2$ synthesized under ambient pressure. As shown in **Fig. 4(a)**, the constituent phases were almost unchanged, and *a*- and *c*-axis length increased and decreased, respectively, as in the case of F-doped to Bi$_2$OS$_2$.

As-synthesized Bi$_2$OS$_2$ did not show large diamagnetism due to superconductivity above 0.47 K. We note that the magnetization curve for as-synthesized Bi$_2$OS$_2$ shown in **Fig. 4(b)** is obtained at 200 Oe, because at low field, dipole like waveform was not obtained at 0.47 K and it was impossible to carry out precise measurement. This apparently indicates that as-synthesized undoped Bi$_2$OS$_2$ is not a superconductor above 0.47 K.

In contrast, HP annealed Bi$_2$OS$_2$ showed superconductivity. The onset of diamagnetic transition is ~4 K, a kink in ZFC curve is at 1.8 K and large diamagnetism is observed below this temperature. HP synthesis was also tried to obtain Bi$_2$OS$_2$ from Bi$_2$O$_3$ and Bi$_2$S$_3$ at 470°C, 2 GPa for 1 h. Bi$_2$OS$_2$ was obtained as the main phase while the amount of impurity phases increased compared to HP annealed samples. Lattice parameters and magnetization of the HP synthesized Bi$_2$OS$_2$ were almost equivalent to those of HP annealed sample. This sample showed small diamagnetism from ~4 K and large diamagnetism is achieved below 2.1 K, as shown in the inset of **Fig. 4(b)**. These results suggest that the $T_c$ of bulk superconductivity for the HP annealed / synthesized samples are ~2 K. The transition at ~4 K is probably due to a impurity phase with small amount such as $Bi_3O_2S_3$, although it is not seen in the XRD pattern.

## 4. Discussion

**Figure 5(a)** shows the $T_c$s at ambient pressure and at ~2 GPa for various $BiS_2$-based superconductors reported to date as a function of $a$-axis length. The values of $a$-axis lengths are measured at room temperature and ambient pressure. $T_c$s are determined by the onset of diamagnetic transition or zero resistivity. In the doped samples, the $T_c$ and $a$-axis length values are those of the optimally-doped ones. $(Sr,La)FBiS_2$ has the longest $a$-axis among these compounds. In $RE(O,F)BiS_2$, $a$-axis shrinks and $T_c$ increases with increasing atomic number of $RE$ from La to Nd. The $a$-axis lengths of $Bi_2(O,F)S_2$ and $Bi_4O_4S_3$ / $Bi_3O_2S_3$, whose blocking layers contain fluorite-type BiO layers, are shorter than that of $Nd(O,F)BiS_2$, although ionic radius (coordination number 6)[27] of Bi is between Nd and Pr. $T_c$s of $BiS_2$-based superconductors under ambient pressure show a dome-like tendency with the top of $T_c$ ~5.5 K at $a$ ~3.98 Å in $(Nd_{0.2}Sm_{0.8})(O_{0.7}F_{0.3})BiS_2$[28]. When $a$-axis in longer than ~4.0 Å, significant increase of $T_c$ is observed in HP resistivity measurement. In contrast, compounds with shorter $a$-axis lengths, $Bi_2(O,F)S_2$ and $Bi_4O_4S_3$, show rapid decrease of $T_c$ in HP resistivity measurement.

The relation between lattice parameter and $T_c$ for as-synthesized and HP annealed $Bi_2(O,F)S_2$ is summarized in **Fig. 6**. In the optimally-doped samples with $a$- and $c$-axes longer than ~3.97 Å and shorter than ~13.73 Å, the value of $T_c$ is maintained at ~5.1 K. In the underdoped samples, $a$-axis is shorter than ~3.97 Å and $c$-axis is longer than ~13.73 Å, and $T_c$ increases as $a$- and $c$-axis expands and shrinks by HP anenaling. $T_c$s for undoped $Bi_2OS_2$ sintered under high pressures are also in this trend. It should be emphasized that in HP annealed / synthesized undoped $Bi_2OS_2$, superconductivity is achieved without intentional carrier doping. In $Bi_2OS_2$, the Bi-S planes are not very flat, the in-plane S-Bi-S angle being 159.8°[25]. The expansion of $a$-axis may lead to flatter Bi-S plane. In $LaOBiS_2$, F-doping not only increases the carrier concentration but also flattens the buckling of the Bi–S plane and this structural transformation is also related to the appearance of superconductivity[29]. Similar phenomena would happen in the undoped and underdoped $Bi_2(O,F)S_2$ by HP annealing, which resulted in the increases of $T_c$s in these samples.

The decrease of $T_c$ in HP resistivity measurement might be explained by the tendency shown in **Fig. 6(a)**. In HP resistivity measurement, $a$-axis might shrink by applying high pressures at low temperatures, and superconductivity could be disappeared. Structural analysis on $Bi_2(O,F)S_2$ under high pressures at low temperatures would provide fruitful information to clear this point.

## 5. Conclusion

High pressure (HP) resistivity measurement and HP annealing were performed for a BiS$_2$-based superconductor Bi$_2$(O,F)S$_2$, which caused different variation of $T_c$. In HP resistivity measurement, $T_c$ linearly decreased at the rate of -1.2 K GPa$^{-1}$. In contrast, by HP annealing at 2 GPa and 470°C, $T_c$ increased in undoped and underdoped samples, and maintained at 5.1 K in optimally-doped sample.

In HP resistivity measurement high pressure is applied *in-situ* at low temperatures, while HP annealing quenches the high pressure and high temperature phase to ambient pressure. Although in both cases hydrostatic high pressure is applied to the sample by a cubic-anvil-type apparatus, the difference between the two methods should be considered carefully. HP annealing technique have been mainly developed on BiS$_2$-based superconductors, but this method can cause unique effects on physical properties of various layered compounds.

**Acknowledgement**

This work was partly supported by the JSPS KAKENHI Grant Number 26390045, and Izumi Science and Technology Foundation. We thank to Dr. T. Fujii and Dr. R. Toda at the Cryogenic Research Center of the University of Tokyo for their cooperation in resistivity measurement at ambient pressure.

*E-mail : tuogino@mail.ecc.u-tokyo.ac.jp

Figure Captions

**Figure 1.** Temperature dependence of resistivity under high pressures for the optimally-doped $Bi_2(O,F)S_2$.

**Figure 2.** $P$-$T$ phase diagram for optimally-doped $Bi_2(O,F)S_2$. $T_c$ for Bi(III) and Bi(V) are cited from Ref. 21 and 22.

**Figure 3.** (a) XRD patterns and (b) Temperature dependence of magnetizations for $Bi_2(O_{1-x}F_x)S_2$ before and after HP annealing

**Figure 4.** (a) XRD patterns and (b) Temperature dependence of magnetizations for undoped $Bi_2OS_2$ sintered at ambient and high pressures

**Figure 5.** $T_c$s on various $BiS_2$-based superconductors at ambient pressure and at ~2 GPa as a function of $a$-axis length at ambient pressure

**Figure 6.** The relationship between lattice parameter and $T_c$ of $Bi_2(O,F)S_2$.

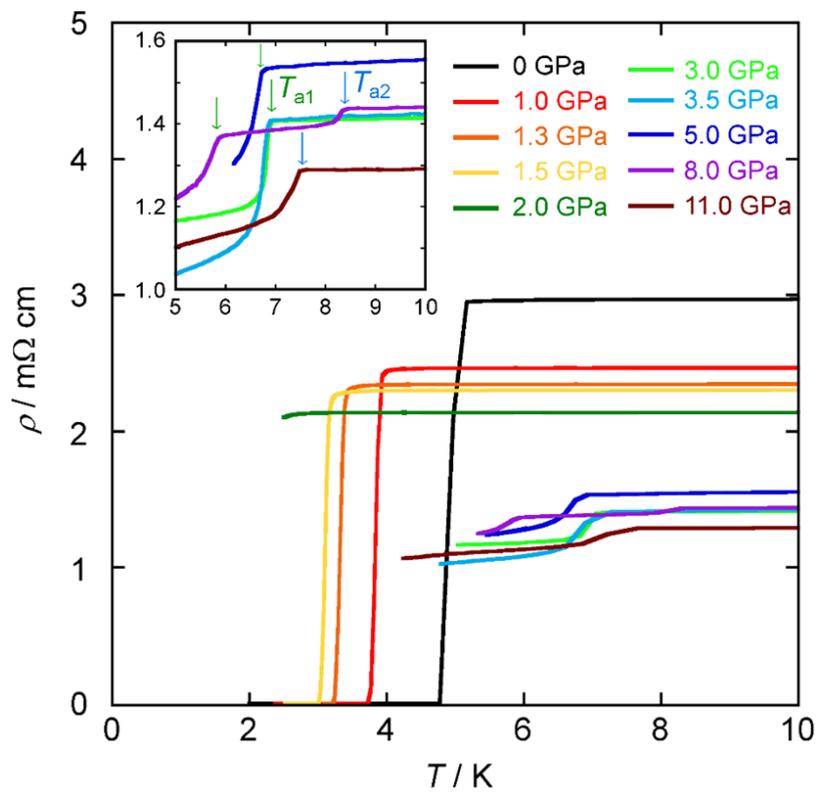

**Figure 1.**

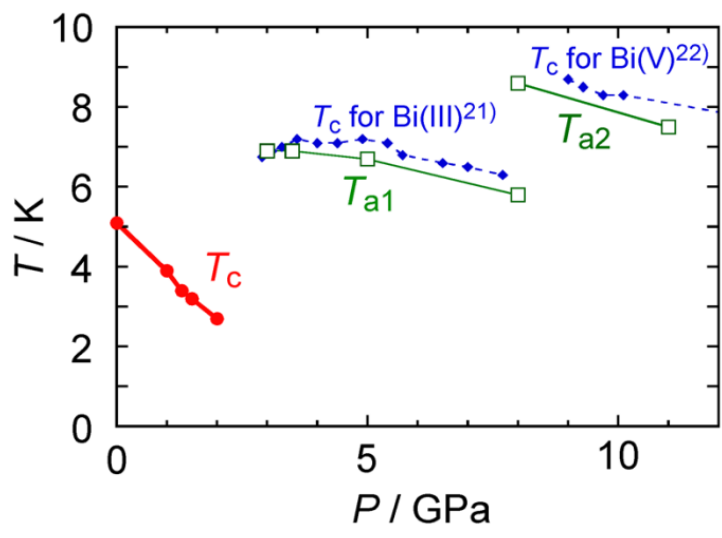

**Figure 2.**

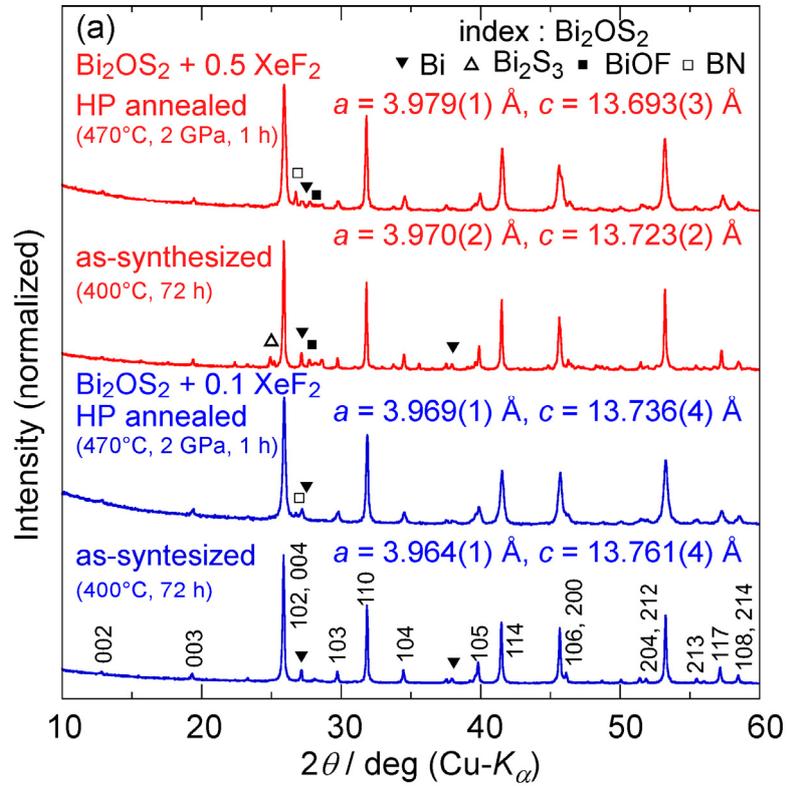

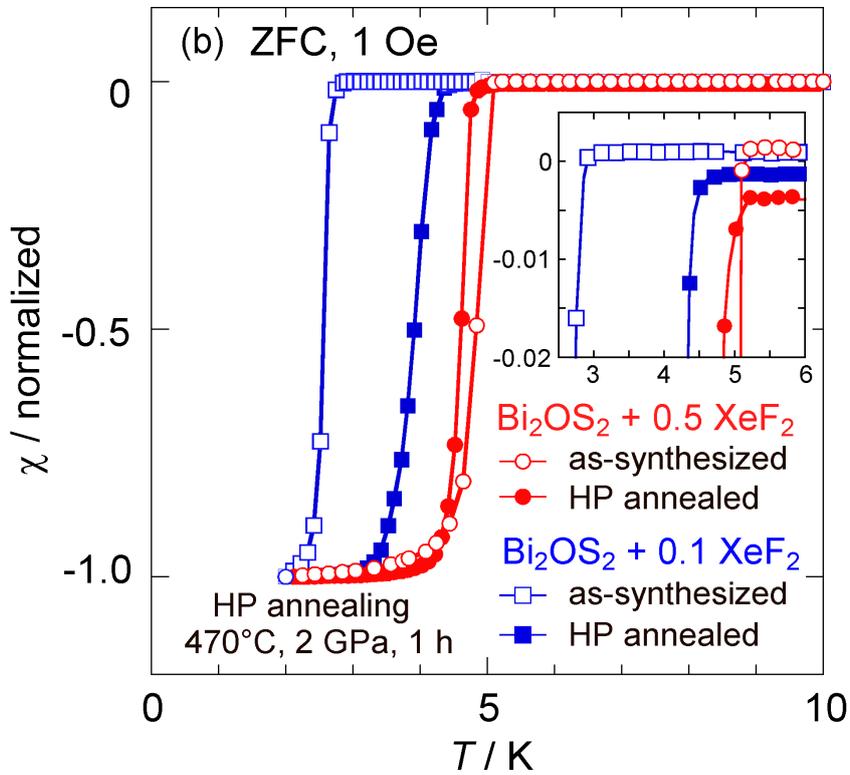

**Figure 3.**

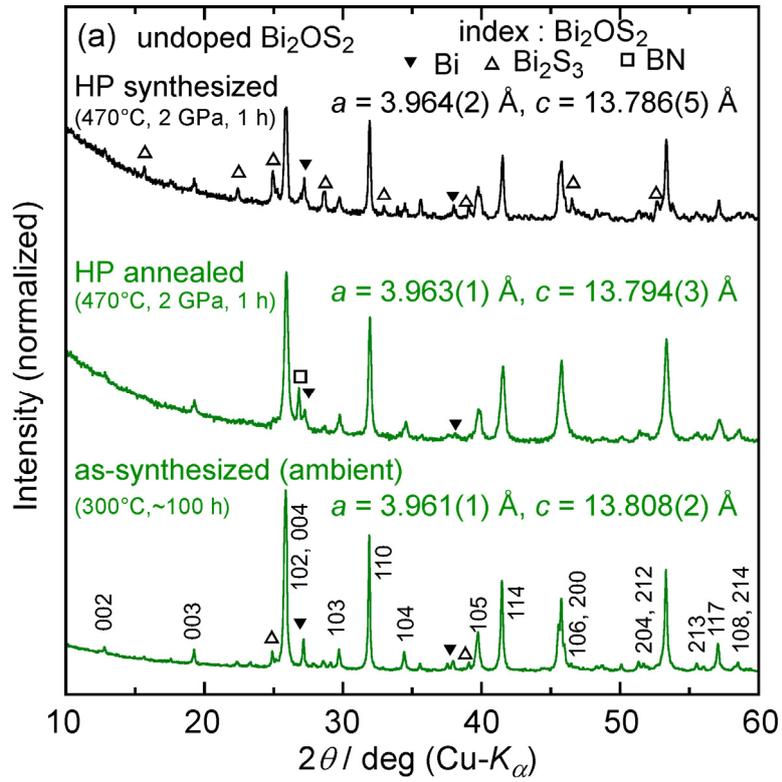
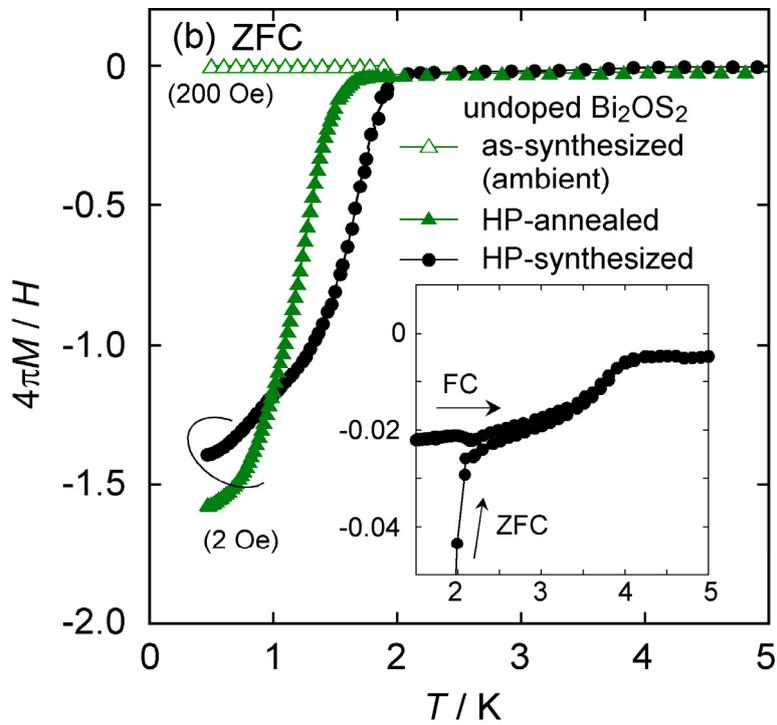

**Figure 4.**

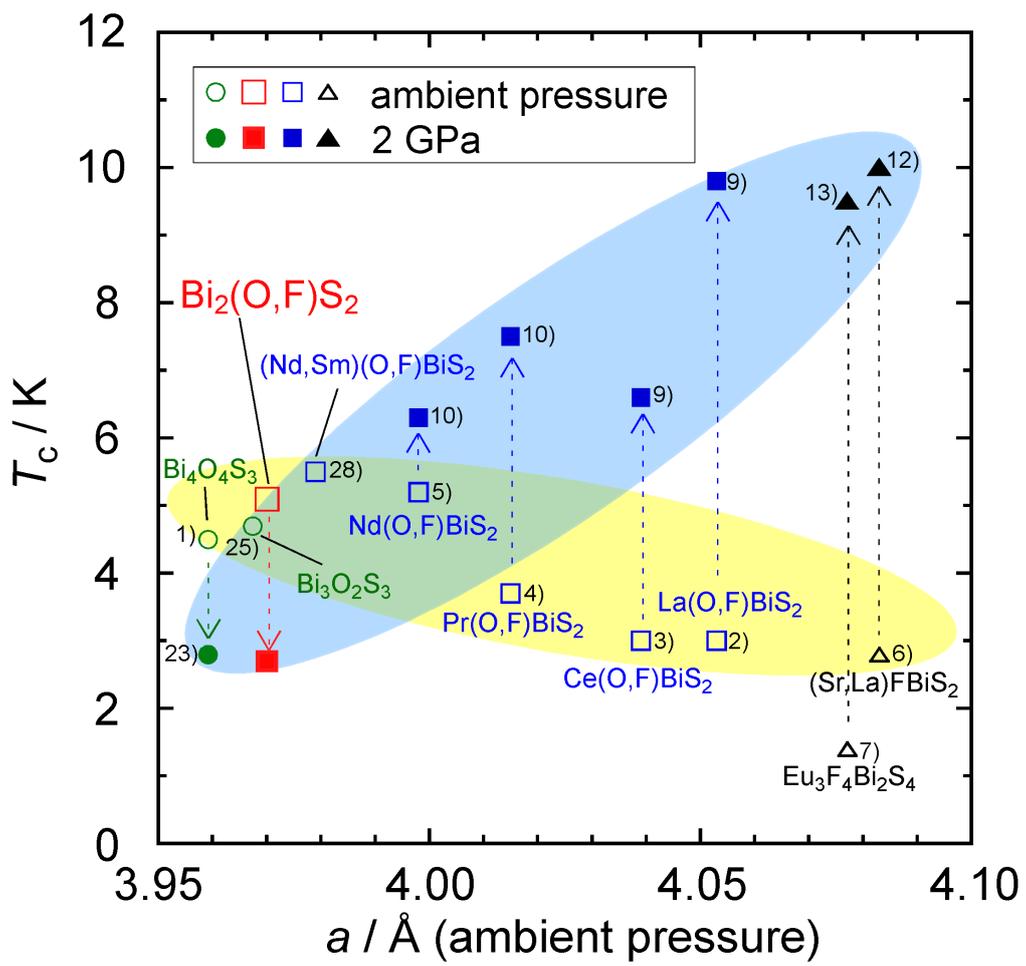

Figure 5.

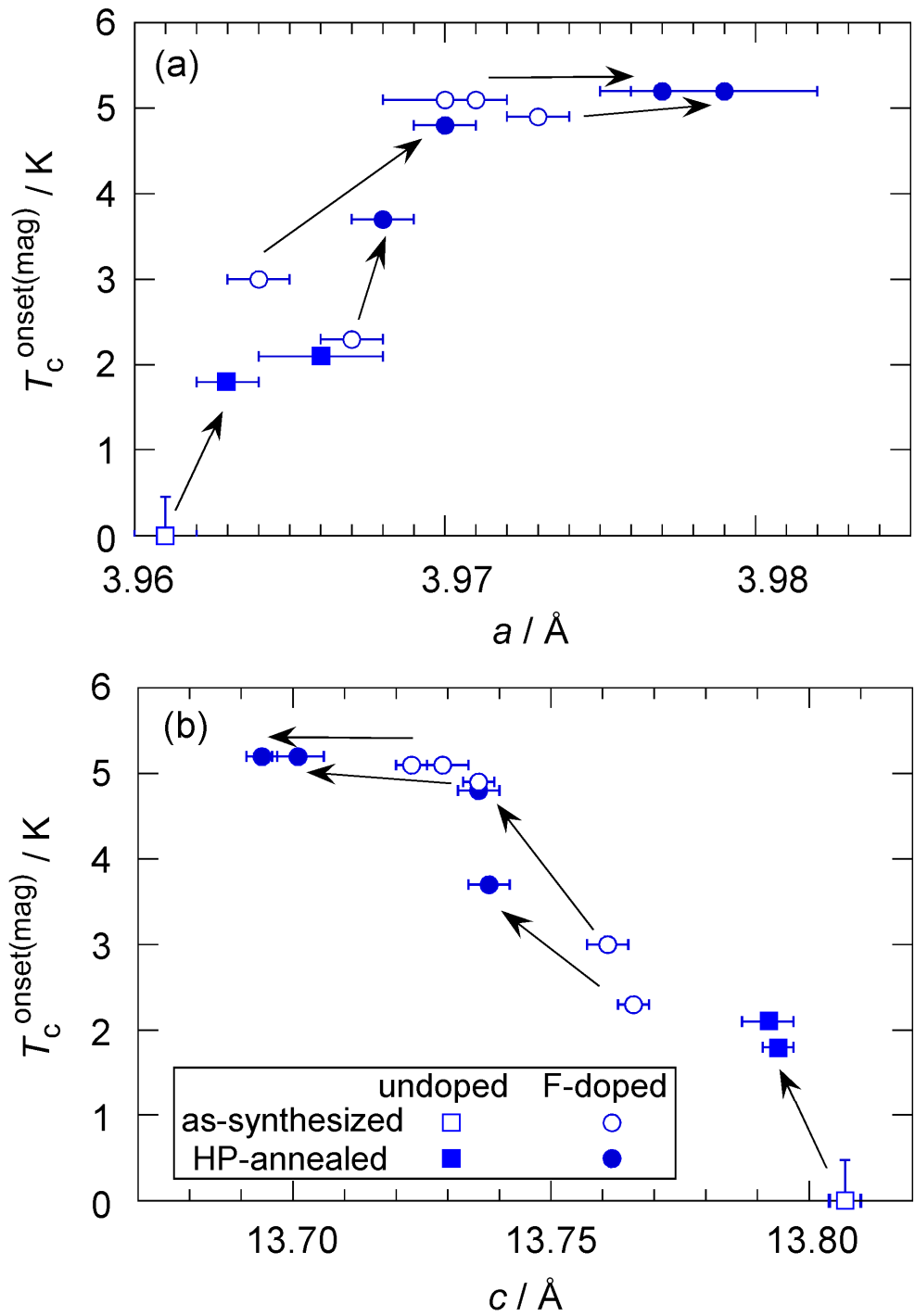

**Figure 6.**